\documentclass[fleqn,usenatbib]{mnras}

\usepackage{newtxtext,newtxmath}

\usepackage[T1]{fontenc}

\DeclareRobustCommand{\VAN}[3]{#2}
\let\VANthebibliography\thebibliography
\def\thebibliography{\DeclareRobustCommand{\VAN}[3]{##3}\VANthebibliography}

\usepackage{graphicx}	
\usepackage{amsmath}	
\usepackage{bm}
\usepackage{ wasysym }
\usepackage{xcolor}
\definecolor{myblue}{rgb}{0.824, 0.89, 0.941}
\definecolor{myblue2}{HTML}{EDF3F9}
\definecolor{myblue3}{rgb}{0.929,0.953,0.976}

\definecolor{myyellow}{rgb}{1, 0.929, 0.8}
\definecolor{mygreen}{rgb}{0.831, 0.925, 0.831}
\definecolor{myred}{rgb}{0.965, 0.827, 0.831}
\definecolor{mygray}{rgb}{0.847, 0.847, 0.847}

\title[Effects of Onset of PT in BNSM]{Effects of Onset of Phase Transition on Binary Neutron Star Mergers}

\author[S. Haque et al.]{
Shamim Haque,$^{1}$\thanks{E-mail: shamims@iiserb.ac.in (SH)}
Ritam Mallick,$^{1}$\thanks{E-mail: mallick@iiserb.ac.in (RM)}
Shashikesh K. Thakur$^{1}$
\\
$^{1}$Department of Physics, Indian Institute of Science Education and Research Bhopal, India
}

\date{Accepted XXX. Received YYY; in original form ZZZ}

\pubyear{2023}

\begin{document}
\label{firstpage}
\pagerange{\pageref{firstpage}--\pageref{lastpage}}
\maketitle

\begin{abstract}

Quantum Chromodynamics predicts phase transition from hadronic matter to quark matter at high density, which is highly probable in astrophysical systems like binary neutron star mergers. To explore the critical density where such phase transition can occur, we performed numerical relativity simulations of binary neutron star mergers with various masses (equal and unequal binaries). We aim to understand the effect of the onset of phase transition on the merger dynamics and gravitational wave spectra. We generated a set of equations of states by agnostically changing the onset of phase transition, having the hadronic matter part and quark matter part fixed.  This particular arrangement of the equation of states explores the scenario of mergers where mixed phases of matter are achieved before or during the merger. Under these circumstances, if the matter properties with hadronic and quark degrees differ significantly, it is reflected in the stability of the final merger product for the intermediate mass binary. We performed a case study on mixed species merger, where one of the binary companions is hybrid star. If quark matter appears at low densities, we observe significant change in post-merger gravitational wave analysis in terms of higher peak frequencies and post-merger frequencies in power spectral density. We report indications expressed as spikes in phase difference plots at merger time for mixed mergers. We found that the expression of phase transition in post-merger gravitational wave signals is more significant for unequal mass binary than for equal mass binary having the same total baryonic mass.
\end{abstract}

\begin{keywords}
neutron star mergers - gravitational waves - dense matter - equation of state - hydrodynamics - methods: numerical
\end{keywords}



\section{Introduction} \label{sec:intro}

\begin{figure*} 
    \hspace{-2cm}
    \begin{minipage}{4cm}
    \includegraphics[scale=0.71]{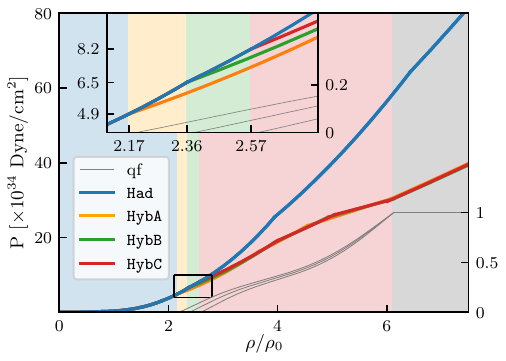}    
    \end{minipage}\hspace{1.95cm}
    \begin{minipage}{4cm}
    \includegraphics[scale=0.71]{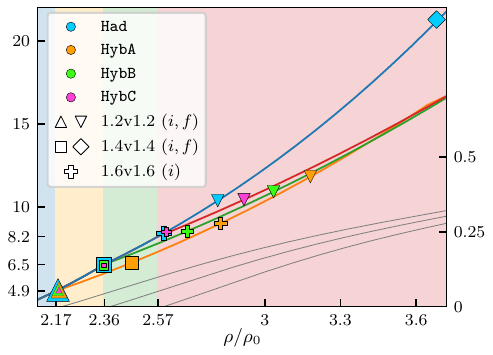}    
    \end{minipage}\hspace{1.8cm}
    \begin{minipage}{4cm}
    \includegraphics[scale=0.71]{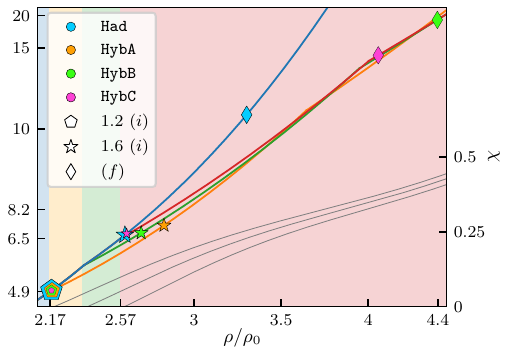}    
    \end{minipage}
    \caption{[Coloured Online] Left: Pressure-density plot of all the EoSs with their respective quark fractions. The inset zooms on the region where EoS splittings are clearly visible, with the tick marks set as per the separation points. Centre: Central (maximum) densities of the initial-($i$) (final-($f$)) stars for all the equal mass binary. Right: Central (maximum) densities of the initial-($i$) (final-($f$)) stars for all the unequal mass binary. The central densities of 1.2 $M_{\odot}$ companion and 1.6 $M_{\odot}$ companion are plotted separately. All the final merger remnant maximum densities are observed after 45 ms post-merger evolution. The marker sizes are adjusted to make all the overlapped markers visible. Differently shaded regions in the background inform the transitions in phases of matter (HM, mixed-phase, pure QM). Light-blue(\textcolor{myblue}{\newmoon}) shaded region is the common density range where all the EoSs have pure HM. Light-yellow(\textcolor{myyellow}{\newmoon}) shaded region (and beyond) shows the presence of mixed-phase in EoS {\ttfamily Hyb-A}. For EoS {\ttfamily Hyb-B}, mixed-phase starts to appear in light-green(\textcolor{mygreen}{\newmoon}) shaded region. EoS {\ttfamily Hyb-C} shows mixed-phase in light-red(\textcolor{myred}{\newmoon}) shaded region. The light-gray(\textcolor{mygray}{\newmoon}) shaded region represents the pure QM regime for all the hybrid EoSs.}
    \label{fig:eos}
\end{figure*}

The advancement of detection capabilities of astrophysical detectors and robust high-resolution numerical simulations of a large number of complicated astrophysical systems has given us the hope to resolve many astrophysics problems, which has intrigued the scientific community for decades. One of the most sought-after questions is about the constituent particles and the fundamental nature of the force that governs matter at high density. Quantum chromodynamics (QCD) predicts a transition (mostly believed as a first order) from hadronic matter (HM) to quark matter (QM) at high density \citep{shuryak}. Although the prediction was made a few decades ago, theoretical or experimental evidence regarding this hypothesis is yet to be found. In fact, constructing an experimental setup to probe such high matter has proved challenging over the years. On the other hand, ab-initio calculations (like lattice QCD~\citep{lattice_qcd}) have failed due to the famous sign problem~\citep{Goy_2017}, and the perturbative QCD calculations~\citep{pqcd} give reliable results only at asymptotically large densities. However, the phase transition (PT) is expected to occur at much lower densities.

Earth-based experiments and theoretical calculations fail in this regard; however, neutron star (NS) astrophysics comes to the forefront. NSs are very dense objects with a mass of few solar masses (most NS mass lies between $1.2\sim1.8~M_\odot$~\citep{Valentim_2011}) and a radius of $\sim10$ km. Therefore, at the centre of such stars, the density is expected to reach a few times that of nuclear density, which is considered an ideal condition where the PT can take place. There had been several works which made an attempt to understand the PT process and its feasibility in isolated NSs \citep{Glendenning_1992,Alford_2001,Alford_2005,Benic_2015,Annala_2020,olinto_1987,Bhattacharyya_2006,Bhattacharyya_2007, Drago_2007, Prasad_2018, Prasad_2020,Baldo_2003,Blaschke_2009,Dexheimer_2010,Orsaria_2014,Ferreira_2020,Han_2020, Prasad_2022, Mallick_2021}. However, verifying such inferences observationally is challenging, as the interior of the NS is not directly visible to us. The detectors can only observe the surface and various emissions from the surface. Presently, the mass of the NS can be measured with high accuracy \citep{Cromartie_2020}, and its radius with improved accuracy up to a few hundreds of meters \citep{Miller_2019,Riley_2019}. Earlier, this was insufficient to constrain the properties of matter (known as the equation of state (EoS)); however, the discovery of a few massive NSs in the last decade has changed the picture entirely.

The discovery of heavy pulsars (highly rotating NSs) like PSR J0348+0432 ($2.01~M_\odot$)~\citep{Antoniodis_2013}, and PSR J0740+6620 ($2.08~M_\odot$)~\citep{Fonseca_2021}, has ruled out soft EoSs which are unable to produce massive NSs. Furthermore, the recent observation of the pulsar PSR J0740+6620 by NICER has given a lower bound on the radius of the pulsar to be $R \gtrsim 11$ km \citep{Miller_2021,Riley_2021}. Such observations have put constraints on the EoS of NS to a great degree. While such observations are gaining momentum, additional constraints came from the observation of binary NS merger (BNSM) GW170817 \citep{GW170817,Radice_2018,Raithel_2019,Baus_2017,Paschalidis_2018,Li_2018,Kashyap_2022}. The gravitational wave (GW) observation of the merger has put a severe constrain of the tidal deformability of the stars~\citep{Hinderer_2010,Most_2018,Annala_2018,Zhang_2018,Nandi_2021}. The present bound on the tidal deformability is $\tilde{\Lambda} \leq 720$. The deformability of the star is directly proportional to the star compactness, which in turn depends on the EoS. Combining all these observations has narrowed down the EoS to a great extent.

Although the observation of the pre-merger phase of GW170817 has helped constrain the EoS to a certain degree, an abundance of mysteries is expected to unfold once we observe the post-merger phase \citep{Sarin_2021}. Numerical Relativity (NR) simulations serve as a promising tool to surf on this domain \citep{Baiotti_2017, Baiotti_2019, Radice_2020, Dietrich_2021,Bauswein_2010_2,Bauswein_2012_2}. There has been a great amount of work which studied the behaviour of BNSM systems and their observables (like GW signals, its Power Spectral Density (PSD) analysis and quasi-universal relations) depending on the EoSs \citep{Raithel_2022_2,Takami_2017, gw_ana,gth,Takami_2014, Jocelyn_2013,Tootle_2021,Sekiguchi_2011_2,Vijayan_2023,Bauswein_2012,Bauswein_2014,Bauswein_2015}, and microphysics (like thermal effects, magnetic field, neutrino radiation, etc.) \citep{Raithel_2022, Most_2022, Most_2020_2, Most_2023, Hanauske_2019,Bruno_2011,Rezzolla_2011,Kenta_2014,Sekiguchi_2011,Kenta_2018,Foucart_2018,Foucart_2020,Blacker_2023,Blacker_2023_2,Just_2023,Henrique_2019}. BNSM systems are rich systems to probe the PT process. There is a significant line of investigations which accounts for quark-hadronic PT (either as smooth crossover or first-order PT) in such systems using GRMHD simulations \citep{Takami_2022,Hanauske_2019_2,Most_2019,Most_2020_3,Weih_2020,Tootle_2022,Kedia_2022,Aviral_2021,Fujimoto_2023}. A line of work involving PT has also been laid using smooth particle hydrodynamics simulations \citep{oechslin_2003,oechslin_2004}, which imposes conformally flat approximation of general relativity \citep{Oechslin_2002}. Some of the studies test the violations of universal relations due to first-order PT in EoSs \citep{Bauswein_2020,Bauswein_2020_2,Bauswein_2019,Blacker_2020}.

It is an accepted fact that at low density, the matter is in the hadronic state. However, the exact density from which QM starts to appear (the deconfinement transition happens) is not known. In this paper, we are simulating BNSM systems (with different mass mergers) to understand the effects of the onset of PT on merger dynamics. We will use a hadronic EoS and the MIT bag model~\citep{mitbag} quark EoS to describe the HM and QM, respectively. These are combined (bridged by Gibbs construction~\citep{gibbs}) to form a hybrid EoS, which is mimicked using piece-wise polytrope to include in simulations. For our set of EoSs, we keep our HM part and QM part of the EoS fixed and only change the onset of PT agnostically. In all the previous works, the QM part differs after the onset point changes, which may happen when an underlying formalism is considered (to include microphysics) to calculate the EoS. Since we used an agnostic approach, we have restricted any such change in our study. From the onset of PT, we smoothly match to the common mixed phase region using the piece-wise polytrope method. Having the HM part and QM part fixed for all the EoSs, we ensure that the difference in BNSM systems appears only due to changes in the onset point. We have studied the possible effects in GW emission spectra and the final merger state (and their lifetimes) depending on the various onset points of PT. It is important to note that this particular choice of the EoSs explores the scenario of mergers where mixed phases of matter are achieved before or during the merger. Secondly, post-merger dynamics are highly affected by other factors like magnetic fields and neutrino cooling \citep{Carlos_2015,Anderson_2008}, which are not considered in this work.

Alongside the set of equal-mass mergers performed, we also reported a case study on a particular unequal-mass merger, which is an NS-HS (neutron star - hybrid star) merger. Here, only one of the companions in the NS pair has a quark seed inside it (an HS); the other companion has no quark seed (an NS). Such mixed systems have been acknowledged in previous work \citep{Bauswein_2020} but require an in-depth study. In this case study, the initial binary configuration is chosen such that the lighter companion is identical for all EoSs. However, the heavier companion has a different interior structure for each EoS. Hence, this case probes the scenario where the effects in the GW emission spectra and the resulting post-merger state are due to the differences in interior initial state properties of the heavier companion only. We have studied how the onset of PT affects these mixed systems. We report the indications from the GW analysis for such a scenario, which are expressed as the spikes in phase difference plots during the merger time. If such an observation is made in the near future, this particular merger will make greater clarity on the understanding of this topic.

The paper is arranged as follows: In Section~\ref{sec:mns}, we explain the formalism and framework used for performing the simulations of the binary merger system and methods used for extracting and post-processing GW signatures. In Section~\ref{sec:results}, we present our results of different equal and unequal mass binary cases (both pre-merger and post-merger simulations). Finally, in Section~\ref{sec:conclusion}, we summarise our results and build conclusions from them.

\section{Methods and Setup} \label{sec:mns}

\subsection{Equation of State}
\begin{figure}
    \centering
    \includegraphics[scale=0.9]{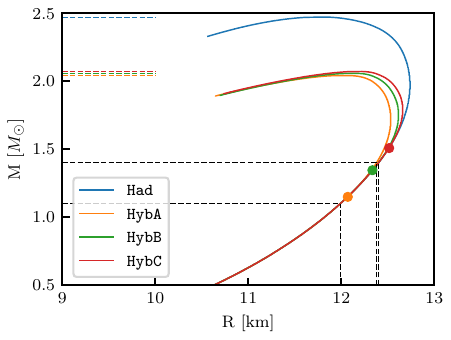}
    \caption{M-R curves for all the EoSs. The dashed lines on the y-axis mark the maximum mass of respective EoSs. The circular markers on the curves define the point after which the heavier NSs will have quark seed. The black dashed lines mark the mass-radius of 1.1 $M_\odot$ and 1.4 $M_\odot$ NSs formed by all the EoSs.}
    \label{fig:mr}
\end{figure}

We used {\ttfamily DD-ME2} EoS~\citep{ddme2} and \texttt{BPS} EoS~\citep{bps1,bps2} to describe the HM and crust respectively. We constructed hybrid EoSs where QM is described using the MIT bag model~\citep{mitbag}, setting the bag constant $B=150$, and parameter $a_4=0.60$. The Gibbs construction is used for a smooth PT of HM to QM~\citep{gibbs}, which creates a {\it mixed-phase} region inside the star where HM and QM coexist. At the point where the mixed-phase starts to appear, we identify that position as the {\it onset point of PT}. We varied the onset points of the mixed-phase in each hybrid EoS by stitching it up at different densities. All the density values are presented in terms of nuclear saturation density ($\rho_0$ = $2.51 \times 10^{14}~\mathrm{g/cm^3}$~\citep{glen}). 
We constructed three hybrid EoSs --- EoS {\ttfamily Hyb-A} (onset point-$2.17\rho_0$), EoS {\ttfamily Hyb-B} (onset point-$2.36\rho_0$) and EoS {\ttfamily Hyb-C} (onset point-$2.57\rho_0$). All the hybrid EoSs have a common point at $6.11\rho_0$ for the transition to pure QM. We mimic the EoSs using piece-wise polytrope fitting~\citep{adm13,ppeos}, plotted in Fig.~\ref{fig:eos} (left). All the EoSs satisfy the following astrophysical constraints --- mass of J0348+0432 ($2.01\pm0.04~M_\odot$)~\citep{Antoniodis_2013}, mass of J0740+6620 ($2.08\pm0.07~M_\odot$)~\citep{Fonseca_2021,Cromartie_2020}, radius measurements by NICER \citep{Riley_2019,Miller_2019,Riley_2021,Miller_2021}. The maximum masses are constrained using J0740+6620 ($2.08\pm0.07~M_\odot$) for all the EoSs, that is, to be greater than $2.01~M_\odot$. Secondly, we use the upper bound on binary tidal deformability (${\tilde\lambda}<720$) by GW170817~\citep{Abbott_2019}. The M-R curves are plotted in Fig~\ref{fig:mr}, with markers on mass-radius of 1.1 and 1.4 $M_\odot$ NSs, and the maximum mass generated by these EoSs. The tidal deformability $\tilde{\Lambda}$ are --- 679, 671, 679, 679 for EoSs --- {\ttfamily Had}, {\ttfamily Hyb-A}, {\ttfamily Hyb-B}, {\ttfamily Hyb-C}, respectively.

These EoSs are supplemented by an ideal-fluid thermal component~\citep{gth2}, which accounts for shock heating in the system that dissipates the kinetic energy into the internal energy. The thermal component $\Gamma_\mathrm{th}$ is set to 1.8~\citep{gth}. More details on the formalism of EoS are available in Appendix~\ref{app:eos}. The constant thermal coefficient does not exactly reproduce the effects of temperature at a high-density regime given by the finite-temperature EoSs \citep{Bauswein_2010,Raithel_2021, Figura_2020}. Secondly, the onset of phase transition could also be affected by the rise in local temperatures. These effects are not considered in the scope of this work. A framework is yet to be found which can create a thermal counterpart from a cold part of the EoS to form a finite-temperature EoS. All available finite-temperature EoSs use a predefined cold part of their own. Hence, for a study like this, where agnostic modifications in the cold EoS are required, it is difficult to use finite-temperature EoS, and hence, the use of a supplemented thermal part becomes necessary.
\begin{figure*} 
\centering\includegraphics[scale=0.41]{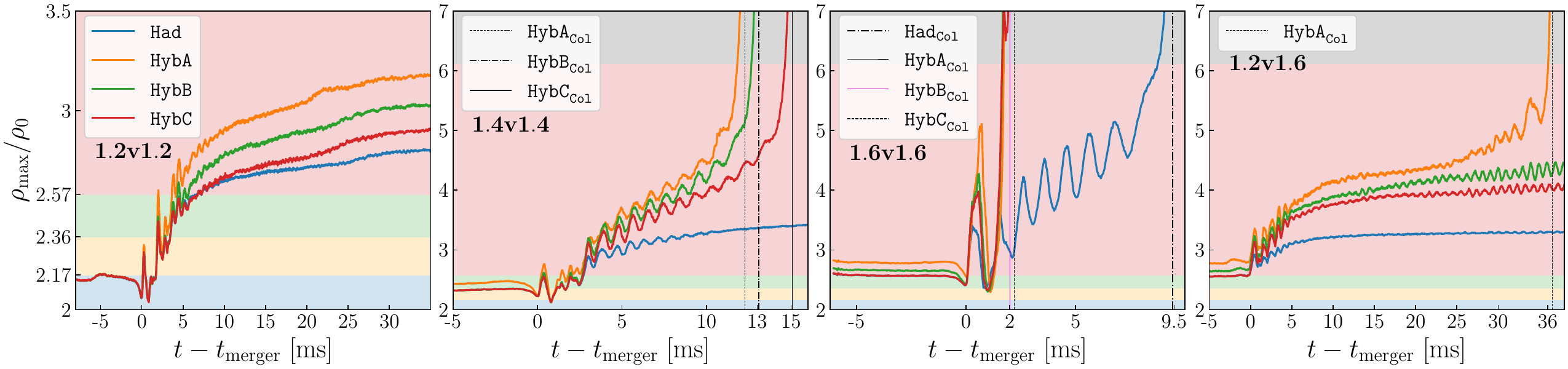}
\caption{\label{fig:rho_max}[Coloured Online] Evolution of $\rho_{\mathrm{max}}$ plotted with mixed-phase and pure QM regions of the EoSs in the background. Collapse times are marked for the cases where compact objects collapsed into BHs.}
\end{figure*}

\subsection{Computational Setup}
For the evolution of the BNSM systems, we use the \textsc{Einstein Toolkit}~\citep{ET2,ETextra1,ETextra2,ETextra3} with \textsc{McLachlan}~\citep{mclachlan1,mclachlan2} (implementation of spacetime evolution) and \textsc{IllinoisGRMHD}~\citep{illinois1,illinois2,p2c1} (GRMHD solver). The initial data are generated using \textsc{Lorene}~\citep{lorene2}. The grid and iteration parameters for our initial configurations were set identical to the version available at the \textsc{Subversion} repository server of the Gravitational Physics Group at the Parma University~\citep{parma1}. We have set the initial physical separation between the stars to be $40~\mathrm{km}$ with the irrotationality of the fluid flow. More details on the numerical setup and uncertainties due to the choice of grid size are available in Appendix~\ref{app:num} and Appendix~\ref{app:grid}, respectively.  We extract gravitational waveforms from our simulations by calculating the Weyl scalar (particularly $\Psi_4$) using the Newman-Penrose formalism~\citep{np}. We analysed the dominant mode $l=m=2$ of ${h}$ strain at 100 Mpc. Now, $h^{22}$ can be given as $h^{22}=h^{22}_+ -ih^{22}_\times = |h^{22}|e^{i\phi}$, where $|h^{22}|$ is the GW amplitude, and $\phi$ is the phase of the complex waveform. We have set the \textit{merger time} at the point where  $|h^{22}|$ is maximum. The \textit{instantaneous frequency} is calculated by $f_\mathrm{GW}=(1/2\pi)(d\phi/dt)$.  We used {{\textsc{Kuibit}~\citep{kuibit1} for handling GW data. 
We calculated the Power Spectral Density (PSD) of the GW amplitude ($\tilde{h}$), given as, $2\tilde{h}^2= |\tilde{h}_+|^2+|\tilde{h}_\times|^2$ where $\tilde{h}_{+,\times}(f)$ are the Fourier transforms of ${h_{+,\times}(t)}$. More details on the formalism of GW analysis are available in Appendix~\ref{app:gw}.

The simulations carried out for this work used the coarsest grid size of $\Delta x = 25 $~$M_\odot$ (with 7 refinement levels), having the finest resolution resolving the star to be $\Delta x = 0.39 $~$M_\odot$ ($\sim 576$~m). The total extent of the 3D domain is $\sim 3\times1000~M_\odot$ ($\sim 3\times1470$~km). This grid setup is considerably low with respect to standard high-resolution simulations carried out in numerical relativity. As discussed in Appendix \ref{app:grid}, the qualitative analysis holds true for respective finer grid simulations. However, the quantitative figures are expected to change if finer grids are used with this framework. Hence, they must be used only as an assistance to the qualitative inferences of this article.

\section{Results} \label{sec:results}

Using the EoS classifications, we simulated three equal mass binary merger cases (1.2v1.2, 1.4v1.4, 1.6v1.6) and one unequal mass binary merger case (1.2v1.6). In some cases, we identify the hypermassive remnant core to be unstable due to high fluctuations of the maximum density in the core. Here, we reported the $\Delta\rho$, which is the local measure of the difference between the highest and lowest values of maximum density in $\sim 1~\mathrm{ms}$. For the scenarios where a hypermaissive remnant core collapses into a Black Hole (BH), it is not straightforward to mark the time stamp when the BH is formed. Hence, calculating the collapse times is non-trivial. In this article, we identify the {\it onset of collapse} when the maximum density in the simulation suddenly peaks ($30 - 100\rho_0$ within $\sim 0.5$ ms) and mark the {\it collapse time} when the maximum density crosses $30\rho_0$. These markers are highly sensitive to additional effects and, hence, must be used for qualitative understanding of the merger dynamics specific to the framework (including grid resolution and EoS set) of this article only. The initial compact stars, which have central densities in the HM region, are identified as NS. If the central density (or maximum) is in the mixed-phase region, we interpret that the star endures a quark seed containing the mixed-phase matter and is identified as an HS. We observe that a lower onset of PT in an EoS leads to a larger quark seed. Similarly, the classifications --- supramassive NS (SMNS), supramassive HS (SMHS), hypermassive NS (HMNS) and hypermassive HS (HMHS) are drawn for the identification of final remnant compact stars \citep{Sarin_2021}. 




\subsection{Equal Mass Binary Simulations}
{Merger-1.2v1.2} --- in Fig.~\ref{fig:eos} (centre), for all EoS cases, we measure the central densities of the initial compact objects to be $2.17\rho_0$ approximately, which is in the pure HM regime. Hence, these initial compact objects are NSs during the inspiral phase. The central densities of NS pair constructed using EoS {\ttfamily Hyb-A} sit at the onset point of PT. In Fig.~\ref{fig:rho_max}, for the EoS {\ttfamily Hyb-A} case, we observe that the mixed-phase appears for a short period due to the spiking maximum density at the merger time. The resulting compact object is an SMHS, as it carries a core ($\rho_{\mathrm{max}}=3.18\rho_0$) with mixed-phase ($\chi=0.23$). As the onset point of PT is set up at higher densities for EoSs --- {\ttfamily Hyb-B} and {\ttfamily Hyb-C}, the simulations introduce mixed-phase in these systems at subsequent times. At the end of the post-merger simulation, the final states are SMHSs with maximum densities of $3.03\rho_0$ ($\chi=0.17$) and $2.91\rho_0$ ($\chi=0.09$) respectively. In all the hybrid EoS cases, the simulations could not reach densities that favour the appearance of pure QM. Secondly, EoS {\ttfamily Had}, being purely hadronic, gives rise to an SMNS with a maximum density of $2.81\rho_0$. GW analysis for this case is reported in Appendix~\ref{app:gw1212}.

Merger-1.4v1.4 --- in Fig.~\ref{fig:eos} (centre), we observe that the compact objects in initial binary configurations made using EoSs --- {\ttfamily Had}, {\ttfamily Hyb-B} and {\ttfamily Hyb-C} are NSs. The NS pair constructed by {\ttfamily Hyb-B} has cores with central densities sitting at the onset of PT. EoS {\ttfamily Hyb-A} has constructed an initial HS binary configuration with comparatively higher central densities for the same initial masses. The cores of these HSs are just inside the mixed-phase regime. In Fig.~\ref{fig:rho_max}, for all the hybrid EoS cases, we observe that the post-merger forms an HMHS, which leads to collapse into a Black Hole (BH). For EoSs --- {\ttfamily Hyb-A}, {\ttfamily Hyb-B} and {\ttfamily Hyb-C}, the collapse times are $\sim$12 ms, $\sim$13 ms and $\sim$15 ms respectively. It infers that the earlier the mixed-phase appears in the system, the faster the HMHS collapses. Hence, the onset point of PT has an important role in defining the collapse times of post-merger remnants. EoS {\ttfamily Had} gives rise to an HMNS, and no collapse scenario is observed up to $45~\mathrm{ms}$ of post-merger evolution. Since the description of mixed-phase in hybrid EoSs is softer than EoS~\texttt{Had}, the systems using hybrid EoSs allow the core of HMHS to crunch more matter. It leads to a quicker increment of core densities, resulting in faster collapse to BH. We infer with certainty that, for this particular case of equal mass binary, hybrid EoSs favour core collapse in contrast to hadronic EoS, which forms a stable HMNS. In a realistic scenario of such mergers, a core collapse can confirm the presence of QM in hypermassive stars and extracting the collapse time can optimise the onset point of PT in our EoS models. 


Merger-1.6v1.6 --- in Fig.~\ref{fig:eos} (centre), we observe that the EoSs --- {\ttfamily Had} and {\ttfamily Hyb-C}, form binary NS configurations. The cores of NS pair constructed using {\ttfamily Hyb-C} are at the onset of PT. The EoSs --- {\ttfamily Hyb-A} and {\ttfamily Hyb-B}, form binary HS configurations. In Fig.~\ref{fig:rho_max}, we observe that all the EoS cases lead to BH collapse scenario. However, the onset of collapse for EoS {\ttfamily Had} case is very far with respect to hybrid EoS cases. For the EoSs --- {\ttfamily Hyb-A}, {\ttfamily Hyb-B} and {\ttfamily Hyb-C}, the rapidly differentially-rotating structure collapses to BH in about 2 ms. In realistic scenarios for high mass binaries, we see that collapse times become an essential parameter to understand the appearance of QM in these systems.

\begin{figure*} 
\centering
\includegraphics[scale=0.88]{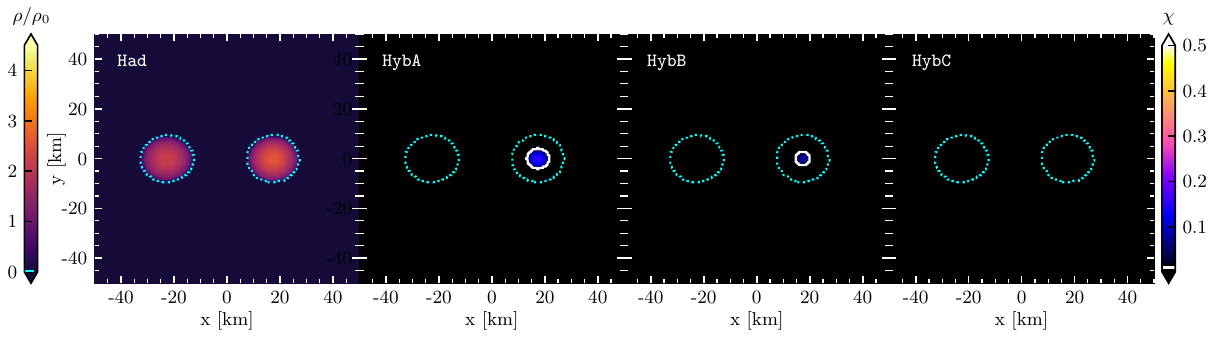}
\caption{\label{fig:ini1216}[Coloured Online] Initial configuration for 1.2v1.6 merger case. Leftmost panel: Initial rest-mass density distribution for EoS \texttt{Had}, contoured with $0.01\rho_0$ (cyan). The remaining plots show the initial quark fraction distribution contoured with $0.01$ (white) and $0.01\rho_0$ (cyan) for EoSs --- {\ttfamily Hyb-A, B, C}.}
\end{figure*}

\begin{figure*} 
\centering\includegraphics[scale=0.83]{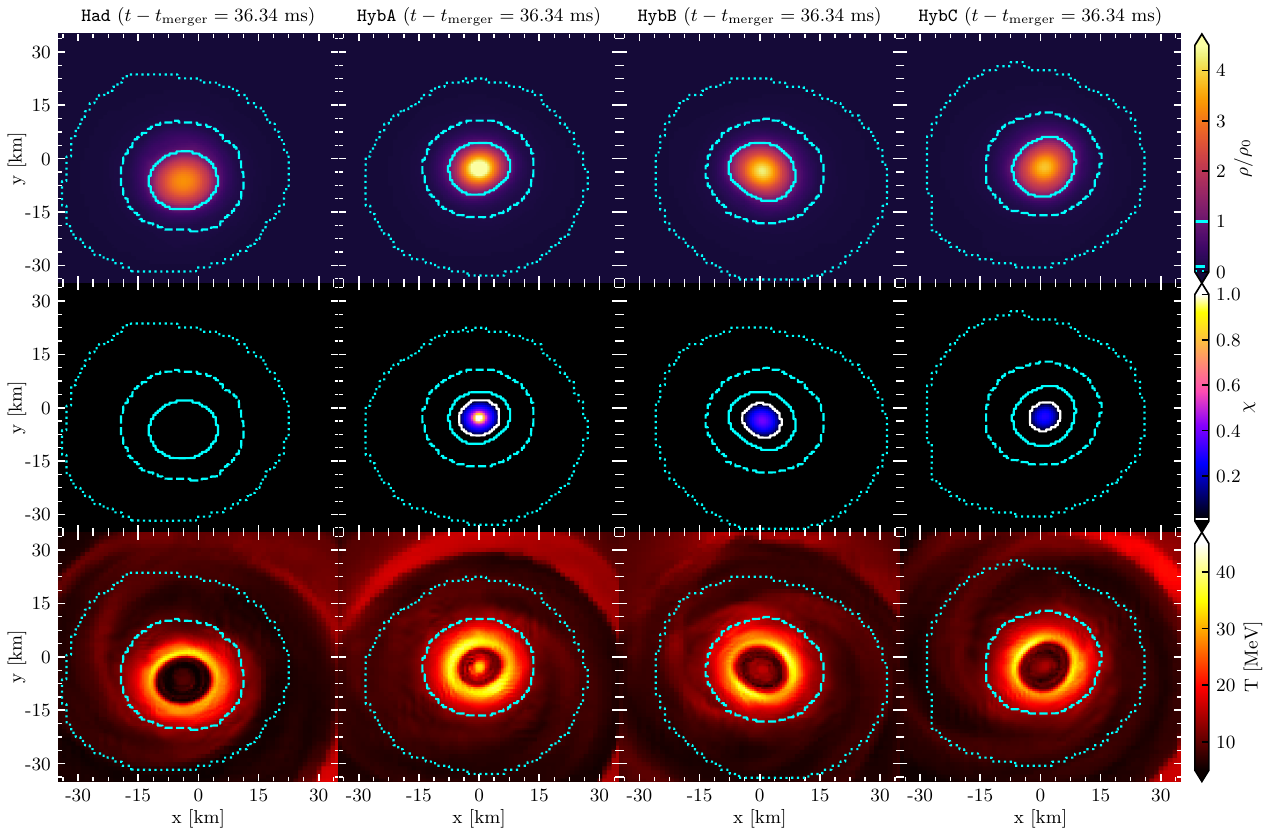}
\caption{\label{fig:fin1216}[Coloured Online] Evolved configuration for all EoSs in 1.2v1.6 merger case at $t-t_\mathrm{merger}=36.34~\mathrm{ms}$. Top panel shows the rest-mass density contoured with $0.01\rho_0$, $0.1\rho_0$, and $\rho_0$ (cyan). Middle panel show the quark fraction distribution contoured with $0.01$ (white), $0.01\rho_0$, $0.1\rho_0$ and $\rho_0$ (cyan). Bottom panel shows the temperature distribution for respective EoS cases contoured with $0.01\rho_0$ and $0.1\rho_0$ (cyan).}
\end{figure*}

\begin{table*}
\caption{\label{tab:12v16} Table of different mass quantities at initial-($i$) configuration (coalescence state) and final-($f$) states (after 36.34~ms of post-merger simulation) for 1.2v1.6 merger case. Column-2 reports the $\rho_\mathrm{max}$ (in terms of $\rho_0$) at the final state. Column-3 reports the initial ADM masses ($M_{\mathrm{ADM}}$) for all the EoS cases. Column-4 and column-5 report the initial and final baryonic masses ($M_{\mathrm{b}}$) of the configurations, respectively. The final $M_{\mathrm{b}}$ is the total baryonic mass calculated inside a 45~km radius sphere. Column-6 reports the quark masses ($M_{\mathrm{q}}$) at the final state. $M_{\mathrm{q}}$ are calculated from uniformly rotating equilibrium configurations of NS (setting the central density and baryonic mass of this NS as $\rho_{\mathrm{max}}(f)$  as $M_{\mathrm{b}}(f)$ respectively) using the RNS code \citep{Stergioulas_1995, Nozawa_1998}. Coulmn-7 reports the quark mass fraction ($X$) at the final state, which is calculated as $M_{\mathrm{q}}/M_{\mathrm{b}}$ at the final state. Column-8 and column-9 report gravitational and baryonic mass at the Keplerian configuration for each EoS, respectively, constructed at $\rho_{\mathrm{max}}(f)$. Column-10 and column-11 report the threshold mass ($M_{\mathrm{thres}}$) \citep{Bauswein_2020_3} for prompt collapse for each EoS at mass ratio $q=1$ and $q=0.7$ respectively.}
\begin{tabular}{ccccccccccc}
\hline
EoS & $\rho_{\mathrm{max}}$ ($\rho_0$) & $M_{\mathrm{ADM}}$ & \multicolumn{2}{c}{$M_{\mathrm{b}}$} & $M_{\mathrm{q}}$ & $X$ & \multicolumn{2}{c}{$M_{\mathrm{Kep}}$} & \multicolumn{2}{c}{$M_{\mathrm{thres}}$}\\
& ($f$) & ($i$) &($i$)&($f$)& ($f$) & ($f$) &($M$)&($M_{\mathrm{b}}$)&($q=1$)&($q=0.7$)\vspace{0.5mm}\\\hline 
{\ttfamily Had}   & 3.29 & 2.48 & 2.80 & 2.65 & 0 & 0 & 2.32 & 2.65 & 3.10 & 3.00\\
{\ttfamily Hyb-A} & 7.32 & 2.48 & 2.80 & 2.63 & 1.38 & 0.52 & 2.36 & 2.72 & 2.84 & 2.67 \\
{\ttfamily Hyb-B} & 4.29 & 2.48 & 2.80 & 2.64 & 0.90 & 0.34 & 2.35 & 2.70 & 2.87 &  2.70 \\
{\ttfamily Hyb-C} & 4.09 & 2.48 & 2.80 & 2.64 & 0.65 & 0.25 & 2.36 & 2.71 & 2.90 &  2.72 \\
\hline
\end{tabular}
\end{table*}

\begin{figure*} 
\centering\includegraphics[scale=0.94]{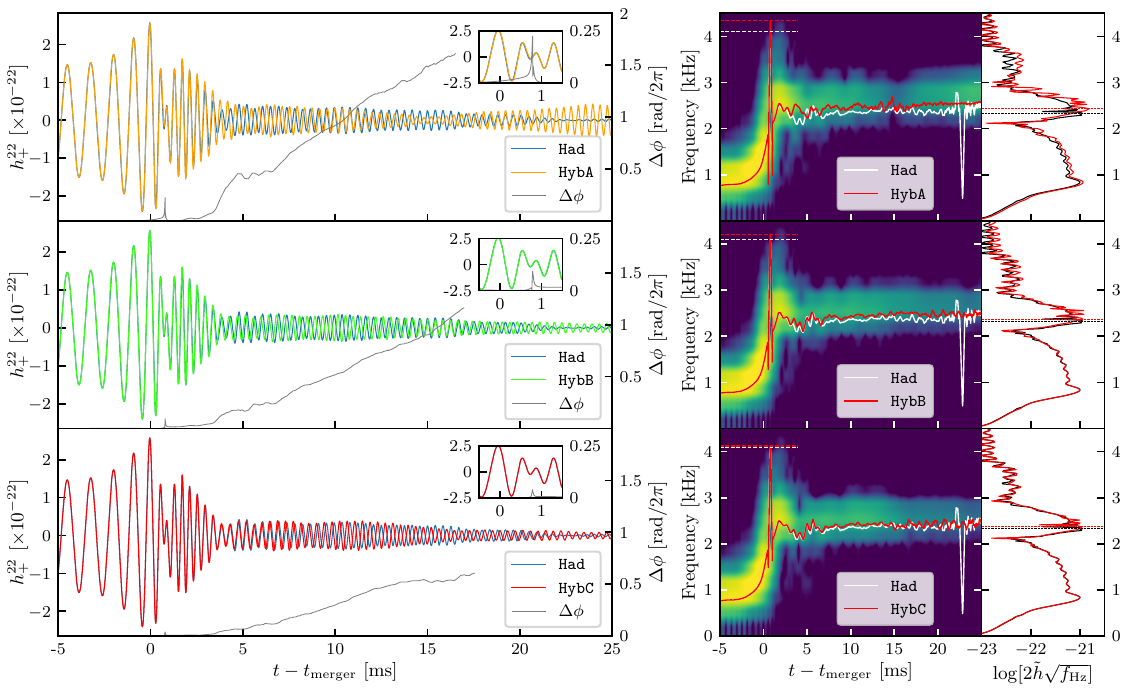}
\caption{\label{fig:1216gw}[Coloured Online] GW emission properties comparison for all the EoSs in 1.2v1.6 merger case. Left panel reports the $h^{22}_+$ polarisation of the GW signal extracted at 100 Mpc. The phase difference $\Delta\phi$ between the two signals is plotted in each case. The inset highlights the spike in $\Delta\phi$ near the merger time. Right panel reports the spectrogram of the GW signal of the hybrid EoS case with the instantaneous frequencies $f_\mathrm{GW}$ of both the signals, marked with peak frequencies. PSDs of both signals are plotted on the right of the spectrogram, marked with the highest power frequencies.}
\end{figure*}

\subsection{Unequal Mass Binary Simulation}

The unequal mass binary (1.2v1.6) case has an interesting contrast to previous equal binary cases. In equal mass mergers, each EoS was constructing compact binaries with distinct interior structures, but the companions in each initial configuration were identical. In the unequal case, all the EoSs have constructed a 1.2 $M_\odot$ (lighter) companion, which is identical for all the initial configurations, but the 1.6 $M_\odot$ (heavier) companion has a different interior structure in each initial configuration. So this case probes the scenario when particularly the heavier companion's interior varies, but the lighter companions remain the same. So, the effects and differences in the GW emission spectra and the resulting state are due to the differences in the initial state properties of the heavier companion. Secondly, an equal mass merger scenario can only allow the mergers of the same species (NS-NS or HS-HS) depending on the combination of initial mass and EoS used for constructing initial data. In this case, our mass choices are such that, by choosing the description of EoS smartly, the combination can allow us to form mixed species (NS-HS) mergers. Such a merger is possible if the lighter companion is purely hadronic and the heavier mass companion has a pre-established mixed-phase core.

\begin{table*}
\caption{\label{tab:sum_all_states}Type of compact objects at initial-($i$) configuration (coalescence state) and final-($f$) states (after 45 ms of post-merger simulation). Respective $\rho_\mathrm{max}$ (in terms of $\rho_0$) are mentioned in brackets with quark fraction in subscript (if HS or HMHS). If the final state is BH, then collapse time (in ms) is mentioned in brackets. For an equal binary case, only one of the star's properties is mentioned because the other companion is identical.}
\begin{tabular}{ccccccccc}
\hline
EoS & \multicolumn{2}{c}{1.2v1.2} & \multicolumn{2}{c}{1.4v1.4} & \multicolumn{2}{c}{1.6v1.6} & \multicolumn{2}{c}{1.2v1.6}\\
&($i$)&($f$)&($i$)&($f$)&($i$)&($f$)&($i$)&($f$)\vspace{0.5mm}\\\hline 
{\ttfamily Had} & NS (2.17) & SMNS ($2.81$) & NS (2.36) & HMNS (3.7) & NS (2.6) & BH ($t_{9}$) & NS-NS (2.18-2.60) & HMNS (3.3) \\
{\ttfamily HybA} & NS (2.17) & SMHS ($3.18_{0.23}$) & HS ($2.47_{0.07}$) & BH ($t_{12}$) & HS ($2.82_{0.16}$) & BH ($t_{2}$) & NS-HS (2.18-$2.82_{0.16}$) & BH ($t_{37}$) \\
{\ttfamily HybB} & NS (2.17) & SMHS ($3.03_{0.17}$) & NS (2.36) & BH ($t_{13}$) & HS ($2.69_{0.08}$) & BH ($t_{2}$) & NS-HS (2.18-$2.69_{0.08}$) & HMHS ($4.4_{0.41}$) \\
{\ttfamily HybC} & NS (2.17) & SMHS ($2.91_{0.09}$) & NS (2.36) & BH ($t_{15}$) & NS (2.6) & BH ($t_{2}$) & NS-NS (2.18-2.61) & HMHS ($4.06_{0.33}$) \\\hline
\end{tabular}
\end{table*}

In Fig.~\ref{fig:eos} (left), we observe that the 1.2 $M_{\odot}$ companion (left) constructed by all the EoS are NSs with central density $2.18\rho_0$ approximately. The central density of NS formed using EoS {\ttfamily Hyb-A} is at the onset of PT. Fig.~\ref{fig:ini1216} confirms that none of these NSs have quark seeds inside them. The 1.6 $M_{\odot}$ companion (right) constructed by EoSs --- {\ttfamily Had} and {\ttfamily Hyb-C}, are NSs with central densities $2.60\rho_0$ and $2.61\rho_0$ respectively, where NS core by {\ttfamily Hyb-C} is at the onset of PT. The 1.6 $M_{\odot}$ companion formed by EoSs --- {\ttfamily Hyb-A} and {\ttfamily Hyb-B}, are HSs, with central densities $2.82\rho_0$ ($\chi=0.16$) and $2.69\rho_0$ ($\chi=0.08$) respectively. Fig.~\ref{fig:ini1216} indicates the same by showing hot spots in the core of these HSs in quark fraction distribution plots. The initial configuration constructed by EoS {\ttfamily Hyb-A} forms an HMHS, which leads to core collapse with a collapse time of $\sim$37 ms (Fig.~\ref{fig:rho_max}). Simulations with EoSs --- {\ttfamily Hyb-B} and {\ttfamily Hyb-C}, led to the formation of HMHSs with unsteady cores having maximum densities of $4.4\rho_0$ ($\chi=0.41$, $\Delta\rho=0.16\rho_0$) and $4.06\rho_0$ ($\chi=0.33$, $\Delta\rho=0.14\rho_0$) respectively. The simulation with EoS {\ttfamily Had} evolves to form an HMNS with a relatively stable core with maximum density $3.3\rho_0$, having negligible fluctuations in the central density. The density, quark fraction, and temperature distributions for the evolved configurations for 1.2v1.6 mergers cases (at $\sim 36~\mathrm{ms}$) are reported in Fig.~\ref{fig:fin1216}. Table \ref{tab:12v16} reports the $\rho_{\mathrm{max}}$ at $\sim 36~\mathrm{ms}$, its initial ADM mass and baryonic mass, final baryonic mass, final quark mass and final quark mass fraction. These quantities are compared with the Keplerian mass and threshold mass for prompt collapse, which are characteristic quantities for each EoS. Such unequal mass mergers make unique observatories for a clear understanding of the onset of PT to mixed phases. Finding such realistic scenarios leading to core collapse can confirm the presence of mixed phases in the systems. Upon being able to reproduce it in simulations, one can further constrain the EoSs, emphasising the onset of PT. 

We compared the GW signal properties of each hybrid EoS with EoS \texttt{Had} for 1.2v1.6 merger scenario in Fig.~\ref{fig:1216gw}. A short yet clear contrast between the GW signals is observed during the merger time (especially for \texttt{Had} v \texttt{Hyb-A} case), which is supported by the spike in $\Delta\phi$ at the merger time. The GW signals deviate at a later stage, as also verified by the increasing $\Delta\phi$. All the inspiral stage systems have mixed-phase seed inside 1.6 $M_\odot$ companion, resulting in the contrast in GW signature. The only exception is the EoS~\texttt{Hyb-C} case, where 1.6 $M_\odot$ NS has no mixed-phase seed (Fig.~\ref{fig:ini1216}). The fact that its central density is at the onset of PT, it still creates enough mixed-phase during the density increase at the time of merging, resulting in a small spike in phase difference. It carries to the consequence that the magnitude of the spike depends on the amount of mixed-phase present at the initial configuration. From the spectrogram, we infer that the peak frequency for EoS~\texttt{Had} is at $\sim 4.1~\mathrm{kHz}$, which is lower than all the hybrid EoSs. The GW signal amplitude flattens after $20~\mathrm{ms}$, making the $f_\mathrm{GW}$ plot noisy in that time regime. For hybrid EoSs --- \texttt{Hyb-A}, \texttt{Hyb-B} and \texttt{Hyb-C}, the peak frequency are $\sim $4.3, $\sim$4.2 and $\sim$4.1 kHz respectively. The peak frequencies are higher for the EoS, which seeds more mixed-phase in the initial configuration. In PSD analysis, we compared the frequencies with the highest power. We found that HMHSs formed in the hybrid EoS cases have higher frequencies than the HMNS in EoS~\texttt{Had} case. Particularly, the HMHS formed in the EoS~\texttt{Hyb-A} case has the highest frequency of $2.4~\mathrm{kHz}$ with respect to EoS~\texttt{Had} with frequency $2.3~\mathrm{kHz}$. It demonstrates that a lower onset point of PT in a hybrid EoS leads to a higher rotational frequency of the hypermassive remnant object.




\section{Conclusions} \label{sec:conclusion}

We simulated equal and unequal mass BNSM systems and observed the change in dynamics when the star is entirely made up of hadrons as to the case when the QM core can appear after the PT. The hybrid EoS has the hadronic degrees of freedom at low density, the mixed-phase region at intermediate density, and pure QM at very high density. The density at which quarks first appear (or the starting point of the mixed-phase region) is the onset point of PT, which is not known priory. To understand the effect of the onset point of PT on the numerical relativity simulations of BNSM systems, we used agnostic approach to construct a set of hybrid EoSs such that the onset of PT is varied, keeping HM and QM parts fixed for all the EoSs. This ensures that the difference in dynamics of BNSM systems appears only due to changes in the onset point. This set of EoSs examines a category of mergers where mixed phases of matter can be achieved before or during the merger.

The initial and final state data of all the simulations are reported in Table~\ref{tab:sum_all_states}. When small mass equal binaries merge, the resultant compact object (Supramassive NS/HS) attains a stable configuration and does not collapse to a BH. In the case of intermediate-mass mergers, if the onset of PT is low, then initial stars may have a quark seed. We observed that the early onset of PT triggered early core collapse into BH. The delay in collapse times hints at the stiffness of the EoS. However, intermediate-mass equal binary constructed with hadronic EoS form an HMNS with a stable core. In the case of heavy-mass mergers, all the mergers collapsed into BH. For the hybrid EoSs, the collapse times did not differ significantly with the change in onset points of PT in the EoS. However, the merger constructed using purely hadronic EoS had notably delayed collapse into BH with respect to hybrid EoS cases.

We reported a case study on a particular unequal-mass merger, an NS-HS merger. These systems are known as mixed merger systems. In this case study, the initial binary configuration is chosen such that the lighter companion is identical for all EoSs. However, the heavier companion has a different interior structure for each EoS. This case probes the scenario where the effects in the GW emission spectra and the resulting post-merger state are due to the differences in interior initial state properties of the heavier companion only. We have studied how the onset of PT affects these mixed systems. We report the indications from the GW analysis for such a scenario, which are expressed as the spikes in phase difference plots during the merger time. Early onset of PT leads to stronger spikes at merger time. If the onset point of PT is at comparatively low density, then the signature of the GW signal is distinct with respect to the GW signal from the hadronic EoS merger, leading to significantly larger phase differences at post-merger. For such a case, we observed higher peak frequencies at the merger time in the GW spectogram. For the same, we observed larger values in PSD frequency with the highest power, indicating that an early onset of PT leads to a higher rotational frequency of the hypermassive remnant. Both the amplitude and frequency of the said post-merger signal are well within the next generation of GW detectors \citep{ET3G}.

The qualitative aspects of the simulation can be improved and made more robust by checking results with an exhaustive set of EoSs and using finer grid resolutions for the numerical relativity simulations (later improves quantitative aspects as well). Since post-merger dynamics are highly affected by phenomena like magnetic fields and neutrino cooling, it is vital to incorporate them into these studies as well. We find that observation of an unequal binary is extremely important to probe into estimating the onset of PT, as the GW signal clearly expresses the change in the onset of PT than in the case of equal mergers. We have seen from the final state of the BNSM that the information about the stiffness of the EoS can be gauged if we can make several observations of such kind. The critical aspects of matter properties, like the degrees of freedom at high density and the information about the onset point, can be extracted when future detectors are available with sensitivity to probe the post-merger phase of the BNSM.

\section*{acknowledgments}
The authors thank IISER Bhopal for providing the supercomputing facility KANAD and other infrastructure facilities to complete this research work. The authors acknowledge National Supercomputing Mission (NSM) for providing computing resources of ‘PARAM Ganga’ at IIT Roorkee, which is implemented by C-DAC and supported by the Ministry of Electronics and Information Technology (MeitY) and Department of Science and Technology (DST), Government of India. SH thanks Z.~B.~Etienne for the help with the \textsc{IllinoisGRMHD}. SH thanks R.~Haas and S.~R.~Brandt for helping to install the Einstein Toolkit on HPCs. SH thanks K.~K.~Nath, D.~Kuzur, S.~Singh, and R.~Prasad for valuable discussions and comments on this work. The authors are grateful to the Science and Engineering Research Board (SERB), Govt. of India for monetary support in the form of Core Research Grant (CRG/2022/000663).This research work is done using {\sc NumPy}~\citep{numpy}, {\sc SciPy}~\citep{scipy}, {\sc Matplotlib}~\citep{matplotlib}, {\sc Mathematica}~\citep{mathematica}, {\sc Kuibit}~\citep{kuibit1}, {\sc Jupyter}~\citep{jupyter}, {\sc VisIt}~\citep{visit}.

\section*{Data Availability}
The data underlying this article will be shared on reasonable request to the corresponding author.



\bibliographystyle{mnras}
\bibliography{main} 




\appendix

\section{Gravitational Wave Analysis for 1.2v1.2 Merger}
\label{app:gw1212}
\begin{figure*}
\includegraphics[scale=0.94]{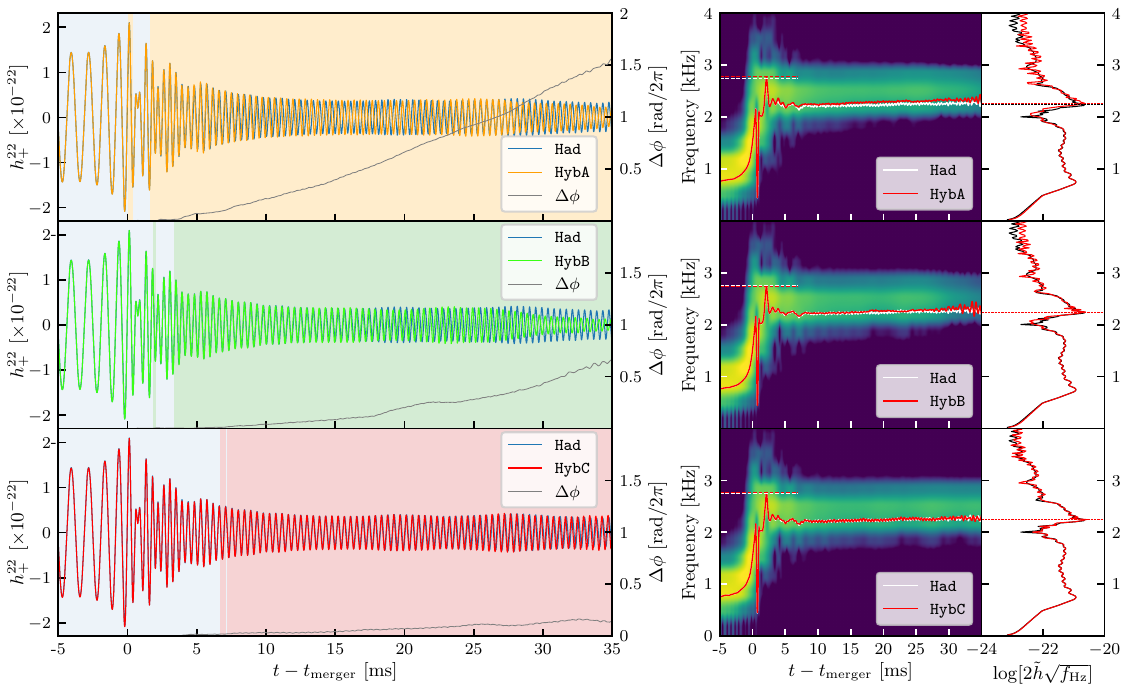}
\caption{\label{fig:1212gw}[Coloured Online] GW emission properties comparison for all the EoSs in 1.2v1.2 merger case. Left panel reports the $h^{22}_+$ polarisation of the GW signal extracted at 100 Mpc for all the hybrid EoSs, overlapped on the GW signal for EoS \texttt{Had}. The phase difference $\Delta\phi$ between the two signals is plotted in each case. The background colours on the plot indicate the presence of pure HM(\textcolor{myblue2}{\newmoon}) matter and the appearance of mixed-phase (respective colours in Fig.~\ref{fig:eos}) in the simulations. Right panel reports the spectrogram of the GW signal extracted from simulations of the hybrid EoS case with the instantaneous frequencies $f_\mathrm{GW}$ of both the signals marked with peak frequencies. PSDs of both signals are plotted on the right of the spectrogram, marked with the highest power frequencies.
}
\end{figure*}

We compared the GW signal properties of each hybrid EoS case with the EoS \texttt{Had} case for 1.2v1.2 merger scenario in Fig.~\ref{fig:1212gw}. Once the mixed-phase appears in the system, the GW signals start to deviate in each case. The divergence can be seen on the overlap and confirmed by the rising phase difference. Since the instantaneous frequencies are not well defined during the merger, $f_\mathrm{GW}$ can peak up or down at the merger time. The second frequency peak is almost equal for all the EoSs and is marked around $2.7~\mathrm{kHz}$. As the GW signals deviate, the hybrid EoS frequencies also start to deviate from EoS \texttt{Had} accordingly. In the case of EoSs \texttt{Hyb-A} and \texttt{Hyb-B}, the simulations tend to gain higher frequencies at post-merger spectra and start to become noisy after $\sim 25~\mathrm{ms}$, that is when the GW signal amplitudes start to decrease. 

In PSD plots, we marked the frequencies with the highest power for each EoS, selectively the one corresponding to the $l=m=2$ fundamental mode of the respective SMNS/SMHSs. They are equal to twice the rotational frequency of the bar deformations of the SMNS/SMHSs. A detailed discussion about the spectral properties of the hypermassive remnants in numerical relativity simulations can be found in Ref.~\cite{gw_ana}. The frequencies are equal for all the cases marked at $2.24~\mathrm{kHz}$ except EoS~\texttt{Hyb-A}, which has a higher frequency marked at $2.27~\mathrm{kHz}$. It indicates that an early appearance of mixed-phase leads to a higher rotational frequency of SMHS. 

\section{Numerical Setup}
\label{app:num}

We used the ``Turing" \texttt{ET\_2021\_05} version of the \textsc{Einstein Toolkit}. It implements the Baumgarte-Shapiro-Shibata-Nakamura-Oohara-Kojima (BSSNOK) formulation via {\sc{McLachlan}}, where $\gamma_{ij}$ is conformally transformed as,
\begin{equation}
\Phi=\frac{1}{12}\mathrm{log}\left(\mathrm{det}\gamma_{ij}\right),\quad\tilde{\gamma}_{ij}=e^{-4\Phi}\gamma_{ij}
\end{equation}
where, $\Phi$ is the logarithmic conformal factor  and $\tilde{\gamma}_{ij}$ is the conformal metric  (constrained by $\mathrm{det}\tilde{\gamma}_{ij}=1$). These are the new variables alongside the trace of extrinsic curvature $\kappa$, the conformal trace free extrinsic curvature $\tilde{A}_{ij}$ and the conformal connection functions $\tilde{\Gamma}^i$, which are defined as,
\begin{equation}
\kappa=g^{ij}\kappa_{ij},~~\tilde{A}_{ij}=e^{-4\Phi}\left(\kappa_{ij}-\frac{1}{3}g_{ij}\kappa\right),~~\tilde{\Gamma}^i=\tilde{\gamma}^{jk}\tilde{\Gamma}^i_{jk}
\end{equation}
\noindent to be evolved using the fourth-order finite-differencing method. The gauge functions are determined using $1+\mathrm{log}$ slicing (for lapse function) and $\Gamma$-driver shift (for shift vectors) condition~\citep{gauge}. During the evolution, a Sommerfeld-type radiative boundary condition~\citep{sommerfeld} is applied to all the components of the evolved BSSNOK variables, and to discard the high-frequency noise, a fifth order Kreiss-Oliger dissipation term is added using the module-\textsc{Dissipation}. 

We used the \textsc{IllinoisGRMHD} code~\citep{illinois1,illinois2} for solving the GRMHD equations, which is an open-source code developed by the Illinois Numerical Relativity (ILNR) group and available with the \textsc{Einstein Toolkit}. The GRMHD equations are defined in a conservative form, and the flux terms are calculated using the second-order finite-volume high-resolution shock capturing (HRSC) scheme~\citep{adm13}. It ensures that Rankine-Hugoniot shock jump conditions are satisfied. A third-order accurate piece-wise parabolic method (PPM)~\citep{ppm} is used for the reconstruction step. The standard Harten-Lax-van Leer-Einfeldt (HLLE) approximate Riemann solver~\citep{hlle1,hlle2} is applied. The method of lines module-\textsc{MoL} takes the time derivatives of the evolved GRMHD variables and integrates them forward in time using the Runge-Kutta fourth-order (RK4) scheme~\citep{rk1,rk2}. A two-dimensional Newton-Raphson solver is employed to compute the primitive variables from the conservative variables~\citep{p2c1,p2c2}. 

\section{Uncertainty of physical quantities due to choice of grid size}
\label{app:grid}

\begin{figure*}
\centering
\begin{minipage}{0.45\textwidth}
\centering\includegraphics[scale=0.45]{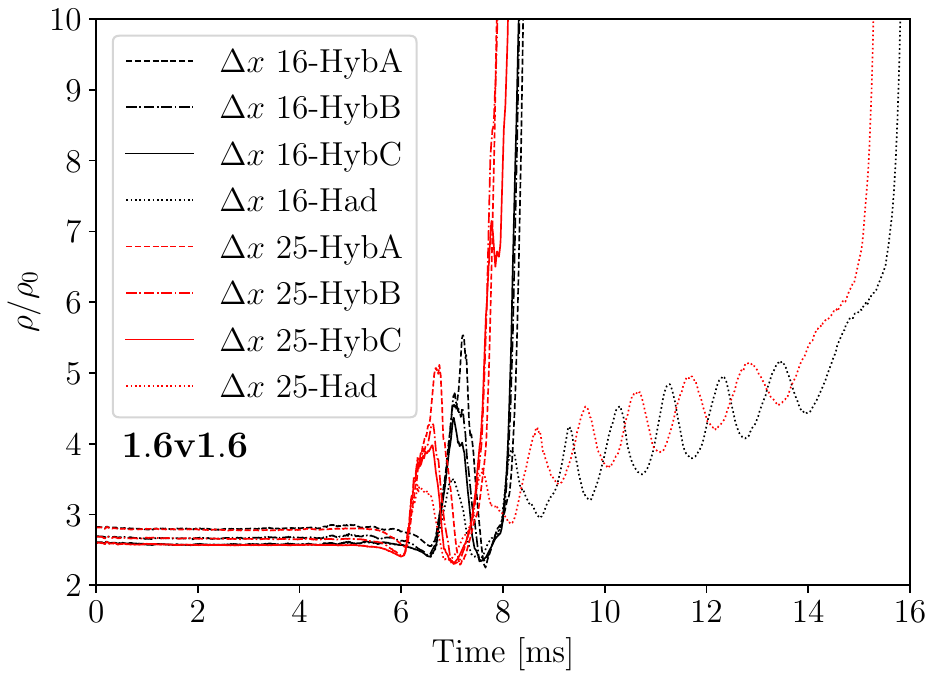}
\end{minipage}
\begin{minipage}{0.45\textwidth}
\centering\includegraphics[scale=0.45]{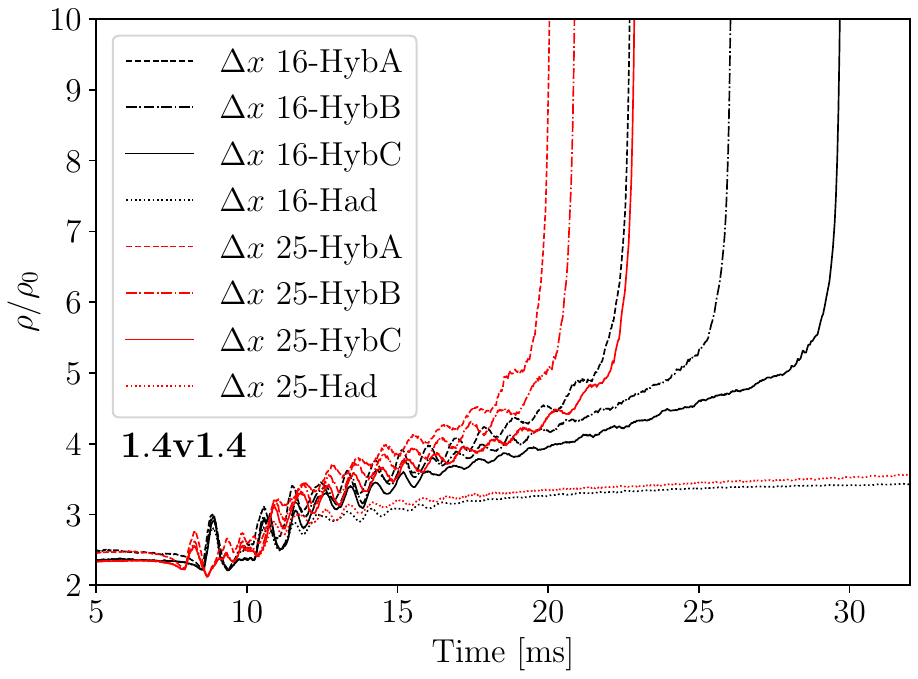}
 \end{minipage}
 \caption{[Left] Comparison of maximum density evolution of 1.6v1.6 $M_\odot$ mergers. [Right] Comparison of maximum density evolution of 1.4v1.4 $M_\odot$ mergers.} 
 \label{fig:rho_16_25}
\end{figure*}

\begin{figure*}
\centering
\begin{minipage}{0.6\textwidth}
\centering\includegraphics[scale=0.45]{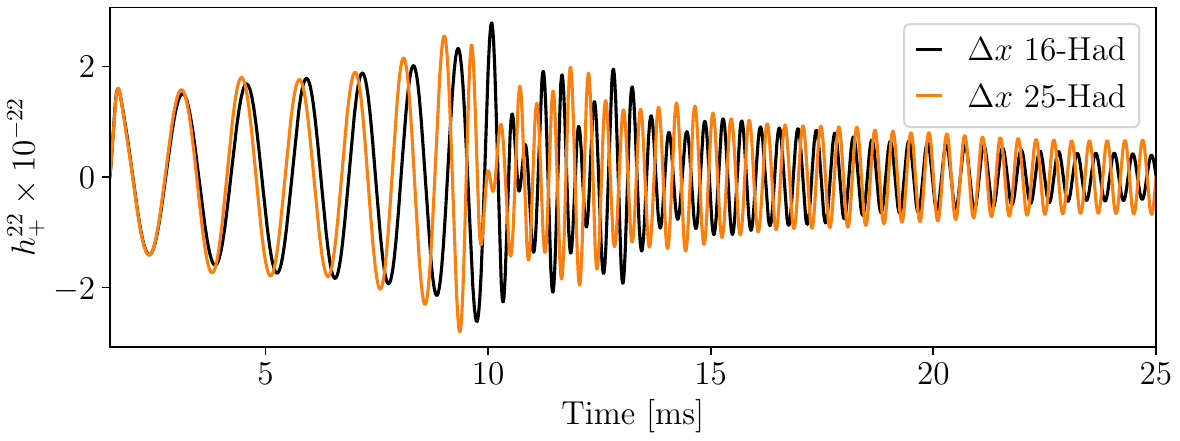}
\end{minipage}
\begin{minipage}{0.39\textwidth}
\centering\includegraphics[scale=0.45]{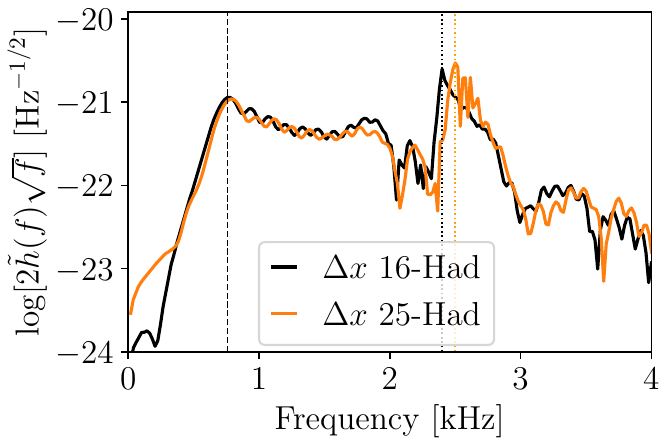}
\end{minipage}

 \caption{[Left] Comparison of the GW strain (observed at 100 Mpc) for simulation of 1.4v1.4 $M_\odot$ merger in grid size $\Delta x=16~M_\odot$ and $\Delta x=25~M_\odot$. [Right] Comparison of the Power Spectral Density (PSD) of the GW strains on left panel.}
  \label{fig:gw_16_25} 
\end{figure*}

The simulations shown in the article have the grid size of $\Delta x = 25 $~$M_\odot$ (with 7 refinement levels) having finest resolution resolving the star to be $\Delta x = 0.39 $~$M_\odot$ ($\sim 576$~m), and the total extent of the 3D domain is $\sim 3\times1000~M_\odot$ ($\sim 3\times1470$~km). 
\newline

To understand the uncertainties of the indicated quantities in the article, we have run a set of simulations with a finer grid size of $\Delta x = 16 $~$M_\odot$ (with 7 refinement levels) having the finest resolution resolving the star to be $\Delta x = 0.25 $~$M_\odot$ ($\sim 369$~m). The summary of this study is described below.

In Fig.~\ref{fig:rho_16_25}, we find a general trend that the simulations with grid size $\Delta x=16~M_\odot$ result in delayed collapse times with respect to the simulations with grid size $\Delta x=25~M_\odot$. However, it is important to note that the qualitative analysis in the article remains unchanged. That is, as the quark matter appears earlier in the matter description, the collapse times are smaller.
    

In Fig.~\ref{fig:gw_16_25}, we find that the GW strains start to differ from the pre-merger phase itself, leading to different merger times. However, we do not observe a change in the corresponding peak (dashed lines at $\sim$ 0.76~kHz coincide) in PSD analysis for the strain. On the other hand, the $f_2$ peaks (dashed lines at $\sim$ 2.5~kHz) differ by $0.1$~kHz.

\section{Equation of State}
\label{app:eos}

The Gibbs condition imposes an equilibrium between HM and QM, keeping the baryonic and electric charges conserved. For the coexistence of two phases, all intensive variables such as pressure, chemical potentials, and temperature of the two phases are identical,
\begin{equation}
    P_{H}\left(\mu^{n}, \mu^{e}, \varphi, T\right)=P_{Q}\left(\mu^{n}, \mu^{e}, T\right)
\end{equation}
\begin{equation}
    \mu^u=\frac{1}{3} \mu^{n}-\frac{2}{3} \mu^{e},\quad
    \mu^{d}=\mu^{s}=\frac{1}{3} \mu^{n}+\frac{1}{3} \mu^{e},
\end{equation}
where $P_H$ and $P_Q$ are HM and QM pressure respectively, $\mu$ is the chemical potential of the species --- neutron ($n$), electron ($e$), up-quark ($u$), down-quark ($d$) and strange-quark ($s$), and $\varphi$ is the solution of hadronic fields. The total baryonic density in the mixed-phase is given as,
\begin{equation}
    \rho_\mathrm{hyb}=(1-\chi) \rho_{H}\left(\mu^{n}, \mu^{e}, T\right)+\chi \rho_{Q}\left(\mu^{n}, \mu^{e}, T\right),
\end{equation}
where $\chi$ is the quark fraction that varies from 0 (pure HM) to 1 (pure QM). At the point where the mixed-phase starts to appear, we identify that position as the ``onset point of PT". The charge neutrality condition is given as,
\begin{equation}
    (1-\chi) q_{H}\left(\mu^{n}, \mu^{e}, T\right)+\chi q_{Q}\left(\mu^{n}, \mu^{e}, T\right)+q_{L}=0,
\end{equation}
where $q_{L}$ is the leptonic charge. The energy density can be written as
\begin{equation}
 \varepsilon_{h y b}=(1-\chi) \varepsilon_{H}\left(\mu^{n}, \mu^{e}, T\right)+\chi \varepsilon_{Q}\left(\mu^{n}, \mu^{e}, T\right).   
\end{equation}We varied the onset points of the mixed-phase in each hybrid EoS by stitching it up at different densities. 


We mimic the EoSs using piece-wise polytrope fitting~\citep{adm13,ppeos}. The process is to first break a tabulated EoS into $N$ pieces (typically $4\sim8$) of density ranges. For each piece $i$ (range $\rho_{i}\leq\rho<\rho_{i+1}$), fit the curve in function,
\begin{equation}
    P_\mathrm{c}=K_i\rho^{\Gamma_i},
\end{equation}
where ($K_i,\Gamma_i$) are the $i^\mathrm{th}$ polytropic constant and polytropic exponent respectively. After fitting a reasonable $\Gamma_i$, the next polytropic constant $K_{i+1}$, is determined using the boundary condition,
\begin{equation}
    K_{i+1}=K_i\rho^{\Gamma_i-\Gamma_{i+1}}_i.
\end{equation} Only $K_1$ is obtained using a reasonable fitting for the first piece. These pieces are matched at the boundary to ensure the smoothness of the mimicked EoS. In piece-wise polytrope fitting of our EoSs, the first three pieces are kept identical for all the EoSs, and the next pieces steer according to further descriptions of matter, respectively. The plots of all the EoSs with respective quark fractions are presented in Fig.~\ref{fig:eos}. The different coloured shaded regions indicate the onset points of PT and mixed-phase regime for respective EoSs, as well as pure HM regime and pure QM regime common for all EoSs.

The EoSs are supplemented by an ideal-fluid thermal component~\citep{gth2}, which accounts for shock heating in the system that dissipates the kinetic energy into the internal energy. It presents the EoS in the form,
\begin{equation}
    P(\rho,\epsilon)=P_\mathrm{c}(\rho)+P_\mathrm{th}(\rho,\epsilon)=K_i{\rho}^{\Gamma_i}+\Gamma_\mathrm{th}\rho(\epsilon-\epsilon_\mathrm{c}),
\end{equation}
where $\epsilon$ is the specific internal energy, and $\epsilon_c$ is given by,
\begin{equation}
    \epsilon_\mathrm{c}=\epsilon_i+\frac{K_i}{\Gamma_i-1}\rho^{\Gamma_i-1}.
\end{equation}
From the following prescription of EoS, the temperature distribution (under ideal-gas approximation) is roughly calculated as \citep{Weih_2020, adm13, Hanauske_2019},
\begin{equation}
    T=\frac{m_nP_\mathrm{th}}{k_B\rho},
\end{equation}
where $m_n$ is the nucleonic mass, and $k_B$ is the Boltzmann constant.


\section{Gravitational Waveform Extraction}
\label{app:gw}

We extract the Weyl scalar (particularly $\Psi_4$) from the simulations using the Newman-Penrose formalism~\citep{np}. Weyl scalars are the projections of the Weyl curvature tensor ($C_{\alpha\beta\mu\nu}$) onto the components of an orthonormal null tetrad ($l^\mu,n^\mu,m^\mu,\bar{m}^\mu$),
\begin{equation}
	\Psi_4=-C_{\alpha\beta\mu\nu}l^\alpha \bar{m}^\beta l^\mu\bar{m}^\nu,
\end{equation}
where Weyl tensor can be given as,
\begin{equation}
	C_{\alpha\beta\mu\nu}=R_{\alpha\beta\mu\nu}-g_{\alpha[\mu}R_{\nu]\beta}+g_{\beta[\mu}R_{\nu]\alpha}+\frac{1}{3}g_{\alpha[\mu}g_{\nu]\beta}R,
\end{equation}
where square braces are the antisymmetric part. This null tetrad is constructed from an orthonormal regular tetrad on each hypersurface based on the spherical coordinate system (bar indicates the complex conjugate). In 3+1 decomposition, by defining a polar orthogonal basis for 3D spatial hypersurface $\left(e^{a}_{\hat{r}},~e^{a}_{\hat{\theta}},~e^{a}_{\hat{\phi}}\right)$ and a normal time-like vector on the hypersurface $e^{a}_{\hat{t}}$, null tetrad can be constructed as,
\begin{equation}
    \small l^a=\frac{1}{\sqrt{2}}\left(e^{a}_{\hat{t}}+e^{a}_{\hat{r}}\right),~k^a=\frac{1}{\sqrt{2}}\left(e^{a}_{\hat{t}}-e^{a}_{\hat{r}}\right),~ m^a=\frac{1}{\sqrt{2}}\left(e^{a}_{\hat{\theta}}+ie^{a}_{\hat{\phi}}\right).
\end{equation}
The complex variable $\psi_4$ provides a measure of outgoing radiation and can be related to the complex GW strain $h$ by its second-order time differentiation~\citep{hdotdot},  
\begin{equation}
    \ddot{{h}}=\ddot{h}_{+}-i\ddot{h}_\times=\Psi_4
\end{equation}

where $h_+$ and $h_\times$ are the polarisation modes. Now, $\Psi_4$ can be can be decomposed using spin-weighted  spherical harmonics~\citep{spin} of spin weight -2,
\begin{equation}
\Psi_4(t,r,\theta,\phi)=\sum_{l=2}^{\infty}\sum_{m=-l}^{l}\Psi^{lm}_4(t,r)_{-2}Y^{lm}(\theta,\phi).
\end{equation} The $\Psi_4$ is provided by the modules-\textsc{ WeylScal4} and \textsc{Multipole}, as a complex grid
function at different coordinate radius surfaces away from the source.


\bsp	
\label{lastpage}
\end{document}